\documentclass[11pt,twoside]{article}

\usepackage{asp2006}
\usepackage{epsf}
\usepackage{psfig}
\usepackage{lscape}

\def\ms{m~s$^{-1}$}
\def\ks{km~s$^{-1}$}
\def\msini{M$_P\sin{i}$}

\def\msun{M$_{\odot}$}
\def\mjup{M$_{\rm Jup}$}

\begin{document}
\title{Planets Around Massive Subgiants}
\author{John Asher Johnson}
\affil{Institute for Astronomy}  

\begin{abstract}
Compared to planets around Sun--like stars, relatively little is known
about the occurrence rate and orbital properties 
of planets around stars more massive than 1.3~\msun. The apparent
deficit of planets  
around massive stars is due to a strong selection bias against
early--type dwarfs in Doppler--based planet searches. One
method to circumvent the difficulties inherent to massive
main--sequence stars is to instead observe them after they have
evolved onto the subgiant branch. We show how the cooler atmospheres
and slower 
rotation velocities of subgiants make them ideal proxies for F-- and
A--type stars. We present the early
results from our planet search that reveal a paucity of planets
orbiting within 1~AU of stars more massive than 1.5~\msun, and
evidence of a rising trend in giant planet occurrence with
stellar mass.  
\end{abstract}

\section{Introduction}

A planet--bearing star can be
thought of as a very bright, extremely dense remnant of a
protoplanetary disk. After all, a star inherits its defining
characteristics---its mass and chemical composition---from the same
disk material that forms its planets. The physical characteristics of
planet host stars therefore provide a crucial link between the planets
we detect today and the circumstellar environments from which they
formed long ago. Studying the relationships between the observed
occurrence rate and orbital properties of planets as a function of
stellar characteristics informs theories of planet formation, and
also helps guide the target selection of future planet searches.

A wealth of recent work has demonstrated that planet occurrence is
strongly correlated with chemical composition \citep{gonzalez97,
  santos04}; metal--rich stars are 3 times more likely to host
planetary companions compared to stars with solar abundances
\citep{fischer05b}. This finding can be
understood in the context of the core 
accretion model. Increasing the metallicity of a star/disk system
increases the surface density of solid material at the disk
midplane, which in turn leads to an enhanced growth rate for
protoplanetary cores \citep{ida04b, kornet05}.

\begin{figure}[ht!]
\plotone{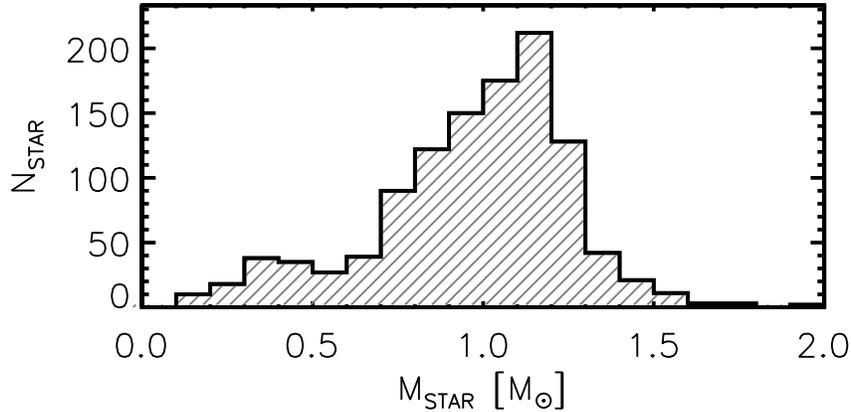}
\caption {\footnotesize{Distribution of stellar masses for the target
  stars of the 
  California and Carnegie Planet Search. The majority of the stars
  have masses between 0.7~\msun\ and 1.3~\msun. \label{mass_hist}}}   
\end{figure}

Another factor that enhances the surface density of material in the
disk midplane is its total mass. If the mass of
circumstellar disks scales with the mass of the central star, then
there should be an observed correlation between planet occurrence and
stellar mass \citep{laughlin04, ida05b, kennedy07}. In principle,
testing this hypothesis is 
fairly simple: one need only measure the fraction of stars with
planets over a wide range of stellar masses. However, in practice such 
a study is not so straight forward given the limited range of stellar
masses encompassed by most planet searches.

The difficulty can be seen in Figure~\ref{mass_hist}, which shows the
distribution 
of stellar masses for the target stars in California and Carnegie
Planet Search \citep[CCPS; ][]{valenti05}, which is representative of
most Doppler--based planet searches. Most of the stars in
Figure~\ref{mass_hist} have masses
between 0.7~\msun\ and 1.3~\msun. In a decidedly
non--Copernican twist of nature, it turns out that stars like our Sun
are ideal planet search targets. Solar--mass G and K dwarfs are
slow rotators, have 
stable atmospheres, and are relatively bright. The fall--off
toward lower stellar masses is simply because late K-- and M--type
dwarfs are faint, making the acquisition of high--resolution spectra
difficult without large telescope apertures \citep{butler06b,
  bonfils05b, endl03}.

The sharp drop at higher stellar masses is due to a separate
observational bias. Stars with
spectral types earlier than F8 tend to have rotationally broadened
spectral features ($V\sin{i} > 50$~\ks; Galland et al. 2005), have
fewer spectral lines due to high surface temperatures, and display a large 
amount of atmospheric ``jitter.'' Stellar jitter is excess velocity
scatter due to surface inhomogeneities and pulsation, which can
approach 50--100~\ms\ for mid--F stars \citep{saar98, wright05}. These 
features conspire to limit the attainable radial velocity precision for
early--type dwarfs to $> 50$~\ms, rendering exceedingly difficult the
detection of all but the most massive and short--period planets.

One method to circumvent the observational limitations inherent to
high--mass dwarfs is to observe these stars after they have
evolved away from the 
main--sequence. After stars have expended their core hydrogen fuel
sources their radii expand, and their atmospheres cool leading to an
increase in the number of metal lines in the star's spectrum. Stars
crossing the subgiant branch also shed a large amount of angular
momentum through the coupling of stellar winds to rotationally
generated magnetic fields \citep{gray85, schrijver93, 
 donascimento00}. The cooler atmospheres and slower rotational
velocities of evolved stars lead to an increased number of narrow
absorption lines in their spectra, making them much better
suited for precision Doppler surveys.

\section{A Doppler Survey of Subgiants}

There are a number of planet searches targeting evolved,
intermediate--mass stars. To date, most surveys have focused on
K--giants \citep{frink02, hatzes05, lovis07, nied07} and ``clump
giants,'' or asymptotic giant branch stars \citep{sato03,
  setiawan03, liu07}. These programs have detected a total of 9 substellar
companions orbiting intermediate--mass giants, demonstrating that
planets do form and can be detected around stars more massive than
$\sim1.5$~\msun \footnote{An additional 5 substellar companions have
  been detected around solar--mass giants}.

Over the past 3 years we have been conducting a Doppler survey
of evolved stars at Lick and Keck Observatories. However,
instead of targeting clump giants and K giants, we have 
focused on stars occupying the region of the H--R
diagram between the main--sequence and red giant branch, known
as the subgiant branch. Our sample is described by \citet{johnson06b}
and is summarized below.

\begin{figure}[ht!]
\plotone{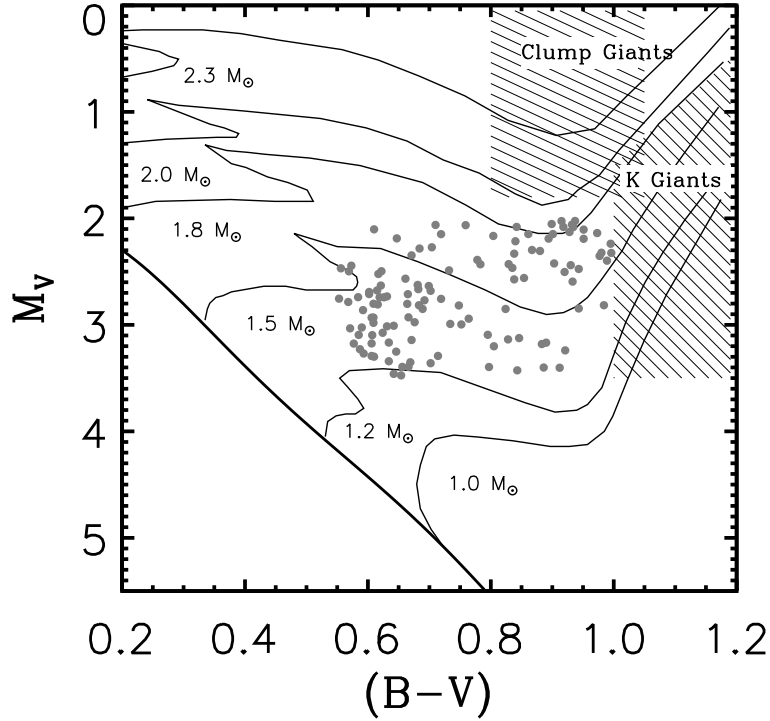}
\caption {H--R diagram illustrating the parameter space spanned by our
  sample of subgiants (circles). Also shown are the regions occupied
  by clump stars and K giants (hashed regions); the theoretical
  stellar mass   tracks of \citet{girardi02} for [Fe/H]$=$0.0
  (thin lines); and the zero--age main sequence (thick diagonal
  line). \footnotesize{\label{sample}}}     
\end{figure}

The main part of our sample is comprised of 120 subgiant stars, which
were selected from the \emph{Hipparcos} catalog based on the criteria $2.0
< M_V < 3.5$, $0.55 < B-V < 1.0$, and $V < 7.6$
\citep{hipp}\footnote{Our full survey contains an additional 39 giant
  stars with $M_V < 2$ and $B-V < 0.85$. However, these stars proved
  to be unsuitable Doppler targets, with a high fraction of close
  binaries and jitter in excess of 30~\ms. We focus here only on the
  more stable subgiants.}. Our sample  
of subgiants is illustrated in an H--R diagram shown in Figure 
\ref{sample}. Also shown are the search domains of other Doppler surveys
containing evolved stars, along with the theoretical mass tracks of
\citet{girardi02}, assuming solar abundances ([Fe/H]=0.0).

Subgiants occupy an observational ``sweet spot'' in the H--R
diagram. They exhibit relatively low levels of jitter, typically
around 5~\ms, which is only a factor of 2 higher than G dwarfs
\citep{wright05} and significantly lower than the 20~\ms\ of jitter
typical for giants \citep{hekker06}. 
Like K giants, they have shed most of their primordial angular
momentum and exhibit slow rotation velocities, with  $V\sin{i} <
5$~\ks. Also, theoretical mass tracks along the subgiant branch
are well separated, allowing for unambiguous mass estimates. Our
sample of stars spans a stellar mass range $1.2 < M_*/M_\odot < 2.2$,
which nearly doubles the stellar mass domain of the CCPS sample.

Our planet search around subgiants has two primary goals. 
First, we wish to compare the orbital characteristics of planets
around intermediate--mass stars to the large statistical ensemble of
planets around lower--mass stars.  Second, we
wish to measure the fraction of stars with planets for stellar masses
$M_* > 1.3$~\msun. To study the relationship between
stellar mass and planet occurrence, we compare the planet fraction
from our high--mass sample to that of the larger sample of FGK stars
in the CCPS and the low--mass M dwarfs from the NASA Keck M
Dwarf Planet Search \citep[e.g.][]{butler06b}. 

\begin{figure}[t!]
\plotone{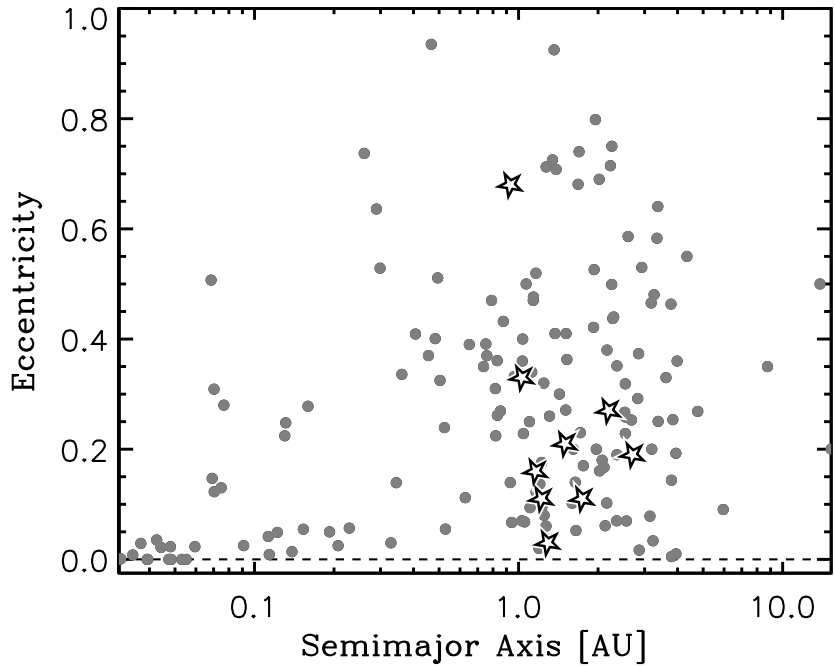}
\caption {Distribution of semimajor axes and eccentricities for
  planets around stars with $M_* < 1.5$~\msun\ (filled circles) and
  subgiants with $M_* > 1.5$~\msun\ (open pentagrams). All of
  the planets around the evolved A--type stars orbit beyond
  $\sim1$~AU. On the other hand, the eccentricity distributions for
  the two sets of planets are comparable. \footnotesize{\label{ae}}}     
\end{figure}

\section{Planet Detections and Characteristics}

The first detection from our sample of subgiants was announced in
\citet{johnson06b}: a short--period, Jovian planet orbiting the
1.28~\msun\ subgiant HD\,185269
\citep[see also][]{moutou06}. With an orbital period $P =
6.838$~d and eccentricity $e = 0.3$, HD\,185269\,b has one of the
highest eccentricities among the sample of known ``hot Jupiters.''
The next batch of planets discovered from our sample orbit stars that
are notably massive: HD\,175541 (1.69~\msun), 
HD\,192699 (1.65~\msun) and HD\,210702 (1.85~\msun, Johnson et
al. 2007a). Following the 
theoretical mass tracks of these three massive
subgiants back to the main sequence reveals that they began life 
as early type dwarfs, with spectral types ranging from A5V to A8V. 

We have recently submitted for publication three additional
long--period planet candidates orbiting the intermediate--mass
subgiants $\kappa$ CrB (=HD\,142091; $M_* = 1.80$~\msun), HD\,167042 (1.65~\msun), and
HD\,16175 (1.35~\msun; Johnson et al. 2007, ApJ submitted). Figure
\ref{ae} shows the distribution of semimajor axes and eccentricities
of all known 
exoplanets orbiting stars with masses $M_* < 1.5$~\msun. Also shown
in the figure are planets orbiting subgiants with masses $M_* >
1.5$~\msun, including the 5 systems announced from our
survey, our two 
strongest unpublished candidates\footnote{Both unannounced systems have false
  alarm probabilities less than 1\%, but lack sufficient phase
  coverage for publication at this time.}, and two other planetary systems
around massive subgiants: HD\,82744 \citep{korzennik00} and HD\,5319
\citep{robinson07}.  

While the eccentricities of
planets around evolved A stars are very similar to those of planets
around Sun--like stars, Figure~\ref{ae} reveals a remarkable trend in
the semimajor axes of planets around high--mass stars.  Planets around
evolved A stars ($M_* > 1.5$~\msun) reside preferentially in wide
orbits at or beyond $\sim1.0$~AU \citep{johnson07}. This observed
semimajor axis 
distribution of planets around high--mass stars differs significantly
from that of planets around lower--mass stars, of which 51\% 
orbit closer than 1~AU. This cannot be due to an observational bias,
since Doppler shift measurements are most senstitive to giant planets
in short--period orbits. While the radii of stars expand as they
evolve away from the main sequence, the radii of subgiants are still
small compared to the semimajor axis of even a $P = 3$~day hot
Jupiter. Thus, it 
remains an open question as to whether the  lack of close--in planets
around A stars is related to the way planets formed in their nascent
high--mass disks, or instead due to the effects of stellar mass on
planet migration.  

\begin{figure}[t!]
\plotone{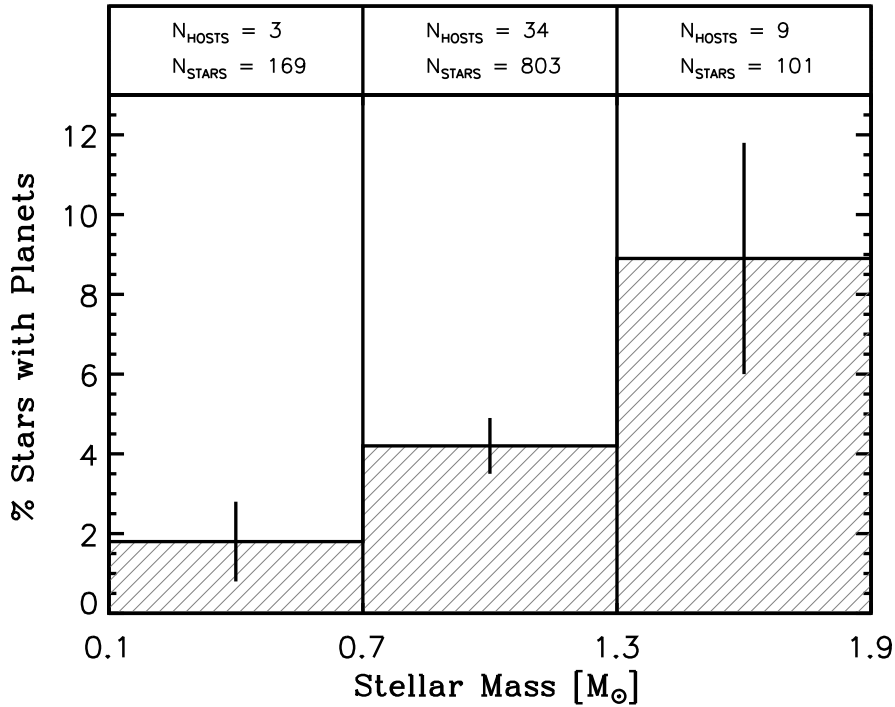}
\caption {The fraction of stars with Jovian planets as a function of
  stellar mass. The error bars represent the uncertainties from
  Poisson statistics. The box above each bin shows the number of
  stars with detected planets, $N_{HOSTS}$, and the total number of target
  stars, $N_{STARS}$. \footnotesize{\label{occ_rate}}}     
\end{figure}

\section{Planet Occurrence vs. Stellar Mass}

Our sample of subgiants covers a range of stellar masses complementary
to the CCPS sample of FGK stars and the sample of
low--mass stars in the NASA Keck M Dwarfs Planet Search. In
\citet{johnson07b} we showed that an analysis of the planet occurrence
rate in three coarsely--spaced mass bins reveals a rising trend with
stellar mass (Figure~\ref{occ_rate}). For
this analysis, we selected target stars and planet candidates with
uniform detection characteristics, namely stars with more than 8
observations, and planets with $a < 2.5$~AU and \msini~$\geq
0.8$~\mjup. 

The observed correlation between stellar mass and planet occurrence
has important implications for planet formation theory, as well as for
future planet search efforts. Stellar mass has now been identified as
an additional sign post of planeticity, along with stellar
metallicity, making A--type stars promising targets for ground--based
imaging surveys as well as space--borne astrometry and transit missions. 

\section{Future Directions}

The study of planets orbiting massive stars is still in its infancy,
with only 17 systems currently known, compared to the 180 Sun--like
and low--mass planet host stars discovered over the past
decade\footnote{For the   updated catalog of extrasolar planet and
  their parameters see http://exoplanets.org.}
\citep{butler06a}. Firmer conclusions about the occurrence rate 
and orbital properties of planets around A--type stars will require a
much larger sample of detections. We have recently expanded our
Doppler 
survey of subgiants to include 300 additional stars at Lick and Keck
Observatories. Our primary goal is to confirm the correlation between
stellar mass and planet occurrence seen in Figure~\ref{occ_rate}. If
the $\sim9$\% occurrence rate holds, we expect to find 20--30 new planets
over the next 3 years. This will represent a significant increase in
the number of planets orbiting evolved A stars, and will allow us to
perform a more robust analysis of the effects of stellar mass on
planetary eccentricities, semimajor axes and multiplicity.

\acknowledgements 
I am very grateful to Geoff Marcy for his inspiration for this
project and his encouragement and advice over the past years. Many
thanks to my collaborators Debra Fischer, Jason Wright, Paul Butler,
Steve Vogt, Chris McCarthy and Katie Peek for their helpful
converstations, and for their work with the data reduction pipelines,
analysis codes and late nights at the telescopes. Thanks to 
Josh Winn and Nader Haghighipour for encouraging me to write up this
summary of my work. I extend my gratitude to the many CAT observers who
have helped  with this project, including Peter Williams, Katie Peek,
Julia Kregenow, Howard Isaacson, Karin Sandstrom,
Bernie Walp, and Shannon Patel. I am an NSF Astronomy and
Astrophysics Postdoctoral Fellow and acknowledge support from the NSF
grant AST-0702821. I also gratefully acknowledge the efforts
and dedication of the staffs of Lick Observatory and Keck Observatory,
and the generous allocation of observing time from the IfA, UCO Lick,
NASA and NOAO TACs.

\end{document}